\documentstyle[12pt,epsf,sprocl]{article}
\def\lsi{\raise0.3ex\hbox{$<$\kern-0.75em\raise-1.1ex\hbox{$\sim$}}}
\def\gsi{\raise0.3ex\hbox{$>$\kern-0.75em\raise-1.1ex\hbox{$\sim$}}}

\newcommand{\la}[1]{\label{#1}}
\newcommand{\ba}{\begin{eqnarray}}
\newcommand{\ea}{\end{eqnarray}}
\newcommand{\rmi}[1]{{\mbox{\scriptsize #1}}}
\newcommand{\fig}{Fig.~}
\newcommand{\fr}[2]{{\frac{#1}{#2}}}
\newcommand{\nr}[1]{(\ref{#1})}
\newcommand{\nn}{\nonumber \\}
\newcommand{\eq}{Eq.~}
\newcommand{\eqs}{Eqs.~}
\newcommand{\tb}{\tan\!\beta}
\newcommand{\be}{\begin{equation}}
\newcommand{\ee}{\end{equation}}

\begin{document}
\begin{flushright}
IEM-FT-205/00\\
IFT-UAM/CSIC-00-30
\end{flushright}
\title{No Spontaneous CP Violation at Finite Temperature in the MSSM?\footnote{To appear in the Proceedings of SEWM2000, Marseille, June 14-17, 2000 }}

\author{P. John}
\address{Instituto de Estructura de la Materia (CSIC), Serrano 123, E-28006 Madrid, Spain\\E-mail: john@makoki.iem.csic.es}

\maketitle

\abstract{In order to generate the baryon asymmetry of the Universe
sufficiently strong CP violation is needed. It was therefore proposed
that at finite temperature there might be spontaneous (transitional)
CP violation within the bubble walls at the electroweak phase
transition in supersymmetric models. We investigate this question in
the MSSM.}
\section{Introduction}
For producing the baryon asymmetry of the Universe, we need extensions
to the Standard Model. One of the Zakharov conditions requires
nonequilibrium. In the MSSM this can be fulfilled by a strong enough
first order phase transition with a light scalar
top\cite{litestop}. Also a small bubble wall velocity\cite{wall}
seems to support baryogenesis. But Zakharov's conditions also require
CP violation and in the MSSM there are several mechanisms known to
generate it.  Explicit CP violating operators might conflict to
experimental EDM bounds\cite{pw}. It were interesting to have a
mechanism generating enough CP violation for baryogenesis without any
conflict to experiments. While spontaneous CP violation is excluded at
$T=0$ for the experimentally allowed parameter values\cite{Pomarol},
there is a suggestion that it might be more easily realized at finite
temperatures\cite{emq,cpr}.

Previously, the moduli of the two Higgs doublets around the phase
boundary have been determined from the 2-loop effective
potential\cite{SecoNum,pj}. The CP violating phase between the two
Higgs doublets has been addressed perturbatively\cite{fkot,John4} and
nonperturbatively\cite{cplr}.

We present the first complete solution of the equations of motion for
the phase between the two Higgs doublets within the MSSM, utilizing a
perturbative effective potential, but without restricting it to the
effective quartic couplings.  Our conclusions\cite{John4} differ from
those obtained earlier on.
\section{Searching for CP violating phases}
We parameterize the two Higgs doublets of the MSSM as  
\be
H_1 = \frac{1}{\sqrt{2}} 
\left(
\begin{array}{l}
h_1 e^{i\theta_1} \\
0
\end{array}
\right), \quad
H_2 = 
\frac{1}{\sqrt{2}}
\left(
\begin{array}{l}
0 \\
h_2 e^{i\theta_2}
\end{array}
\right). \label{Hparam}
\ee
In addition, because of gauge invariance, the effective Higgs
potential depends on the phases only via $\theta=\theta_1+\theta_2$,
and we have an additional constraint $h_1^2 \partial_\mu \theta_1
=h_2^2\partial_\mu \theta_2$.  We can then concentrate on
$\theta$. Assuming tree-level kinetic terms and moving to a frame
where the bubble wall is static and planar, the action to be minimized
is
\be
S \propto \int dz \Bigl[
\fr12 (\partial_z h_1)^2 + 
\fr12 (\partial_z h_2)^2 + 
\fr12\fr{h_1^2h_2^2}{h_1^2 +h_2^2}(\partial_z \theta)^2 + V_T(h_1,h_2,\theta)
\Bigr],\label{action}
\ee
where $V_T(h_1,h_2,\theta)$ is the finite temperature effective
potential for $h_1,h_2,\theta$.  In general, we are solving the
equations of motion for $h_1,h_2,\theta$ following from this
action. In the numerical solution 
we use the method outlined in~\cite{pj} which 
deals with the minimization of a functional of 
the squared equations of motion.

At the first stage, we consider the case with no explicit CP phases, 
and ask whether 
a particular solution without CP violation ($\theta=0,\pi$), 
is in fact a local minimum of the action or not. Clearly, it
is not if
\be
m_3^2(h_1,h_2)\equiv
\frac{1}{|h_1h_2|}\left.\frac{\partial^2V_T(h_1,h_2,\theta)}
{\partial\theta^2}\right|_{\theta=0}<0, \label{constraint}
\ee
where we have divided by $|h_1h_2|$, 
assuming that this is non-zero. 
\eq\nr{constraint} is to be evaluated along the path found 
by  solving the equations of motion for $h_1,h_2$.
We have chosen the convention that $h_1$ can have either sign,
allowing us to consider only $\theta=0$.
For the case of the most general quartic
two Higgs doublet potential, \eq\nr{constraint} agrees 
with the constraint on which most of the investigations of 
spontaneous CP violation are based. However, 
\eq\nr{constraint} is true more generally, independent of the 
form of the potential $V_T(h_1,h_2,\theta)$. 

The tree-level potential of the theory is
\be
V_\rmi{tree}=  
\frac{1}{2}m_1^2 h_1^2 + \frac{1}{2}m_2^2 h_2^2 + 
m_{12}^2 h_1 h_2\cos\theta 
+\frac{1}{32}(g^2+{g^\prime}^2)( h_1^2- h_2^2)^2, \la{tree}
\ee
where $g,g'$ are the SU(2) and U(1) gauge couplings, and
at tree-level
\be
m_{12}^2 = -\frac{1}{2}m_A^2\sin2\beta. \la{m12}
\ee 
It follows that $m_3^2(h_1,h_2) = (1/2) m_A^2 \sin\!2\beta > 0$, 
so that the minimum of the potential in the 
$\theta$ direction is at $\theta=0$. Thus, in order
to get spontaneous CP violation one needs radiative 
corrections which can overcome the tree-level term. 

Older considerations\cite{emq,cpr,fkot} are based on the approximation to
the effective potential where only the quadratic and quartic operators
are considered. At finite temperatures around the electroweak phase
transition, important contributions come from infrared sensitive
non-analytic contributions which are not of this form, and can affect
spontaneous CP violation\cite{cplr,John4}.  Thus, it is important to
solve the equations of motion more generally for the full effective
potential.

Here we consider the full finite temperature 1-loop
effective potential of the MSSM. It is known that 2-loop corrections
are very important in the MSSM\cite{litestop}, allowing for larger
values of $h_1,h_2$ in the broken phase. Nevertheless, for the present
problem we find that even 1-loop effects are in most cases very small,
so we do not expect qualitative changes from the 2-loop effects.
\section{A scan for spontaneous transitional CP violation.}

The tree-level part of the effective potential $V_T(h_1,h_2,\theta)$
is in~\eq\nr{tree}.  In the resummed 1-loop contribution to
$V_T(h_1,h_2,\theta)$, we include gauge bosons, stops, charginos and
neutralinos. This introduces dependences on the trilinear squark
mixing parameters $A_t$ and $\mu$ as well as on the squark mass
parameters $m_Q^2$, $m_U^2$, and the U(1), SU(2) gaugino parameters
$M_1$ and $M_2$.

We now wish to see whether the constraint in \eq\nr{constraint}
can be satisfied at the bubble wall between the symmetric
and broken phases. To do so, we have to search for each parameter
set for the critical temperature $T_c$, solve the equations
of motion for $(h_1,h_2)$ between the minima, and evaluate
$m_3^2(h_1,h_2)$ along this path. Since this is 
quite time-consuming, we proceed in two steps.

{\bf 1.} 
At the first stage, we do not solve for $h_1,h_2,T_c$, but rather
take them as free parameters in the ranges 
$h_1/T=-2..2$ and $h_2/T=0..2$, $T=80...120$ GeV.
The zero temperature parameters are varied in the wide ranges
\ba
& & \tan\beta  = 2...20, \qquad 
m_A = 0...400 \mbox{ GeV}, \nn
& & m_U =  -50...800 \mbox{ GeV}, \qquad
m_Q = 50...800 \mbox{ GeV}, \\
& & \mu, A_t, M_1, M_2 = -800...800  \mbox{ GeV}. \nonumber
\ea
Here a negative $m_U$ means in fact a negative 
right-handed stop mass parameter, $-|m_U^2|$. We have also
studied separately the (dangerous\cite{cms}) region where 
the transition is very strong\cite{litestop,cms},
corresponding to $m_U \sim -70...-50 \mbox{ GeV}$. 

Note that since we
do not solve for the equations of motion at this stage but 
allow for $h_1=\pm |h_1|$, we have to divide
in \eq\nr{constraint} by $h_1h_2$ instead of $|h_1h_2|$:
this leads in general to positive values due to the 
tree-level form of the potential, \eqs\nr{tree},\nr{m12}. 
A signal of a potentially promising region is then 
a small absolute value of the result, since this means 
that we are close to a point where $\partial_\theta^2 V_T(h_1,h_2,\theta)$
crosses zero. 

{\bf 2.}
At the second stage, we study the most favourable
parameter region thus
found in more detail. First of all, we search for the critical
temperature. Then, we solve the equations of motion for $(h_1,h_2)$. 
By comparing with the exact numerical solution in several cases, 
we find that a sufficient accuracy can be obtained in practice
by searching for the ``ridge'' as an approximation to the wall
profile. It is determined as the line of maxima 
of the potential in the direction
perpendicular to the straight line between the minima. 
Finally, we look for the minimum of $V_T(h_1,h_2,\theta)$
at fixed $(h_1,h_2)$: this is a fast and reliable approximation
for the full solution in the case that $\theta$ is small
(i.e., just starts to deviate from zero), and 
corresponds to \eq\nr{constraint}.

For the first stage, we perform a Monte Carlo scan with about $2\cdot
10^9$ configurations.  Small values of $m_3^2(h_1,h_2)$ are scarce,
and even then do not necessarily correspond to the desired phenomenon
of spontaneous CP violation: they could also be points far from the
actual wall. This can be clarified at stage 2.

\begin{figure}[t]
\epsfxsize=12pc
\vspace*{-0.5cm}
\centerline{\epsfbox{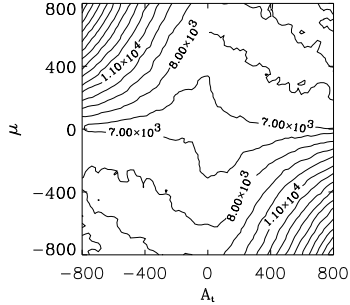}} 
\vspace*{-0.5cm}
\caption{The average value of $m_3^2$ versus $\mu$ and $A_t$. 
We observe that small values of $m_3^2$ are not typical in any part
of the plane but are on the average more likely for small $\mu,A_t$,
and that the distribution is wider (and thus more favourable) for
like signs of $\mu,A_t$, as shown by the noisy contours obtained
with a finite amount of statistics.}
\label{Atmu}
\end{figure}

The parameter region found depends most strongly on $m_A$, $\tb$, with
a preference on small values of $m_A$ and large of $\tb$, such that
$m_{12}^2$ in \eq\nr{m12} is small. (This is in contrast to the
requirements of a strong phase transition\cite{litestop}). There is
also a relatively strong dependence on $A_t$ and $\mu$: the region
favoured is shown in \fig\ref{Atmu}. The dependences on the other
parameters are less significant; for $m_U$ and $m_Q$ small values are
preferred. The region found is in rough agreement with those found
in~\cite{emq,cpr,fkot,cplr}.

At the second stage, we make further restrictions.  For instance, we
exclude the cases leading to non-physical negative mass parameters. We
also exclude cases leading to $T=0$ spontaneous CP violation in the broken
phase: this phenomenon requires very small values\cite{Pomarol} of
$m_A$. We also discard phase transitions which are exceedingly weak,
$v/T\ll 0.1$.

In\cite{fkot} the special point $m_U^2\approx 0$ was
considered. Since in\cite{fkot} the thermal mass corrections were
neglected this corresponds in the physical MSSM to a case where $m_U^2
+ \#T^2\sim 0$.  Expanding the 1-loop cubic term from the stops to a
finite order in $v_1/v_2$, it was suggested that transitional
spontaneous CP violation can take place. This region is quite
dangerous due to the vicinity of a charge and colour breaking
minimum\cite{cms}. Without expanding the 1-loop contribution in
$v_1/v_2$, we cannot reproduce the behaviour proposed there.  In any
case, even before taking into account the experimental lower limits on
the Higgs masses, we cannot find any promising case in the sample of
$\sim 2\times 10^6$ configurations of stage 2.

We conclude\cite{John4} that after taking into account the infrared
sensitive effects inherent in the 1-loop effective potential, coming
from a light stop and gauge bosons, and solving for the wall profile
from the equations of motion, spontaneous CP violation does not take
place in the physical MSSM bubble wall.

\noindent{\bf Acknowledgements} I would like to thank S.~Huber, M.~Laine, M.~Schmidt for collaboration in~\cite{John4}.

\end{document}